\documentclass[nofootinbib,twocolumn,showpacs]{revtex4}
\usepackage{amsmath}
\usepackage{amsfonts}
\usepackage{epsfig}
\def\argmax{\mathop{\rm argmax}}

\def\lad{\lambda^{\downarrow}}

\def\<{\langle}
\def\>{\rangle}
\def\n{\noindent}
\newtheorem{definition}{Definition}
\newtheorem{theorem}{Theorem}
\newcommand{\beq}{\begin{equation}}
\newcommand{\eeq}{\end{equation}}
\newcommand{\beqa}{\begin{eqnarray}}
\newcommand{\eeqa}{\end{eqnarray}}
\newcommand{\bra}[1]{\ensuremath{{\langle{#1}|}}}
\newcommand{\ket}[1]{\ensuremath{{|{#1}\rangle}}}

\begin{document}

\title{Optimal Control of Quantum Dissipative Dynamics: Analytic
solution for cooling the three level $\Lambda$ system}
\author{Shlomo E. Sklarz}
\author{David J. Tannor}
\affiliation{Department of Chemical Physics, Weizmann Institute of
Science, Rehovot, Israel 76100. Tel 972-8-9343723, Fax 972-8-9344123}
\author{Navin Khaneja}
\affiliation{Division of Engineering and Applied Sciences, Harvard
University, Cambridge MA, USA}

\begin{abstract}
We study the problem of optimal control of dissipative quantum
dynamics. Although under most circumstances dissipation leads to
an increase in entropy (or a decrease in purity) of the system,
there is an important class of problems for which dissipation with
external control can decrease the entropy (or increase the purity)
of the system.  An important example is laser cooling. In such
systems, there is an interplay of the Hamiltonian part of the
dynamics, which is controllable and the dissipative part of the
dynamics, which is uncontrollable.  The strategy is to control the
Hamiltonian portion of the evolution in such a way that the
dissipation causes the purity of the system to increase rather
than decrease. The goal of this paper is to find the strategy that leads 
to maximal purity at the final time. 
Under the assumption that Hamiltonian control is complete and 
arbitrarily fast, we provide a general framework by which to calculate
optimal cooling strategies. These assumptions lead to a great simplification,
in which the control problem can be reformulated in terms of the spectrum of 
eigenvalues of $\rho$, rather than $\rho$ itself. By combining this
formulation with the Hamilton-Jacobi-Bellman theorem we are able to obtain
an equation for the globaly optimal cooling strategy in terms of the spectrum
of the density matrix.
For the three-level $\Lambda$ system, we provide a complete analytic
solution for the  optimal cooling strategy.
For this system it is found that the optimal strategy does not exploit
system coherences and is  a 'greedy' strategy, in which 
the purity is increased maximally at each instant.
\end{abstract}
\maketitle
\section{Introduction}
In the last 15 years, optimal control theory (OCT) has been
applied to an increasingly wide number of problems in physics and
chemistry whose dynamics are governed by the time-dependent
Schr\"odinger equation (TDSE).  These problems include control of
chemical reactions
\cite{Tannor85,Tannor86,Tannor88,Kosloff89,Rice2000,Shapiro03,Brixner03,Mitric02},
state-to-state population transfer
\cite{Peirce88,Shi91,Jakubetz90,Shen94}, shaped wavepackets
\cite{Yan93}, NMR spin dynamics \cite{Khaneja01}, Bose-Einstein
condensation \cite{Hornung,Sklarz02.1,sklarz02.2}, quantum
computing \cite{Rangan01,Tesch01,Palao02}, oriented rotational
wavepackets \cite{Leibscher03}, etc. \cite{Rabitz,Gordon97}. More
recently, there has been vigorous effort in studying the control
of systems governed by the Liouville-von Neumann (LVN)  equation,
where the central object is the density matrix, rather than the
wavefunction
\cite{Bartana93,Bartana,TannorRot,Cao97,Gross98,Ohtsuki03,Khaneja03}.
The Liouville-von Neumann equation is an extension of the TDSE that
allows for the inclusion of dissipative processes. Important examples
of what may be thought of as quantum control processes that require
the use of the LVN include laser control of chemical reactions in
solution, laser cooling, and coherence transfer in multi-spin systems.
In all these cases, the external field (the laser or the RF field) is
the coherent control, while the source of dissipation is contact with
the environment.  In the case of laser cooling, the environment is the
vacuum modes of the electromagnetic field and the source of
dissipation is spontaneous emission.

In the majority of problems on control of quantum systems dissipation
is a nuisance; the purpose of the control is to either avoid,
delay or cancel the dissipation process.  Yet there is a
remarkable exception to this pattern --- laser cooling. The goal
of laser cooling is expressed alternatively as increasing the
phase space density, or decreasing the entropy of the system.
Purely Hamiltonian manipulations can in fact do neither, and
therefore dissipation, rather than being a nuisance, is actually
necessary to achieve true cooling. The optimal control of systems
of this type is fascinating.  The control itself, no matter what
its time-dependence, leads only to Hamiltonian evolution and hence
no true progress toward the objective.  On the other hand, the
dissipation, while it is capable of producing progress toward the
objective, is fundamentally not controllable and could in fact
lead to a decrease in the objective.

In ref. \cite{TannorRot}, we elucidated the interplay of the
controlled, Hamiltonian evolution, and the uncontrolled,
dissipative evolution in producing cooling.  The ``cooling laser'',
while not directly cooling the system, in fact steers it to a
region of parameter space where spontaneous emission leads to
cooling rather than heating. We define such a controlled
manipulation as a "purity increasing transformation".   We believe
that the study of such transformations in their general
mathematical context is of extreme interest, both in terms of
discovering a wider class of physical processes where purity, and
therefore coherence content can be increased, as well as because
of the rich mathematical structure of the problems involving
interplay of Hamiltonian and dissipative dynamics.

In \cite{TannorRot}, we solved the problem of optimal cooling for
a 2-level system completely, under the assumption of complete and
rapid Hamiltonian control. We showed that the optimal cooling
strategy in the 2-level system avoids producing coherences in the
density matrix. Here we present a general framework for the analysis
of optimal control in a system of $N$ excited states coupled
radiatively to $M$ ground states, under the same assumptions.
Using this framework we explicitly provide the optimal strategy for 
cooling of a three level $\Lambda$ system.

We first introduce the Lindblad dissipation model and a
generalized concept of purity in section \ref{sec:Prelim}. In
Section \ref{sec:Prob} the problem of optimal cooling of a
quantum mechanical system is formulated. It is shown in Section 
\ref{sec:Reform} that this
problem can be reformulated solely in terms of the eigenvalue
distribution of the density operator. In doing this, we derive a
reduced equation of motion for the spectral evolution under
dissipation, parameterized by the unitary control (Sections
\ref{sec:Motion} and \ref{sec:Canon}). Section \ref{sec:HJB}
introduces the mathematical tools for finding optimal cooling
strategies, namely the Hamilton-Jacobi-Bellman theorem. Section 
\ref{sec:3lvlAppl} provides an explicit description of the optimal cooling
strategy for the three level $\Lambda$ system and proves its
optimality. Finally we discuss future directions and conclude in
Section \ref{sec:Conc}.

\section{\label{sec:Prelim}Setting up the control problem}
\subsection{\label{sec:LVN}The system equations of motion and the Lindblad formula for dissipation}
\begin{figure}[ht]
\centerline{\epsfig{file=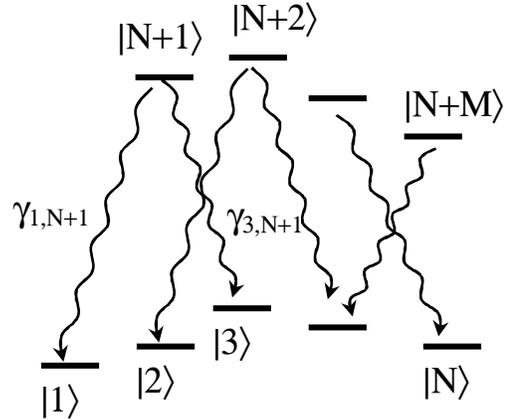,clip=width=2.5in}}
\caption[N+Mlvlsys]{A general $N+M$ level quantum system with
spontaneous emission rates between various energy levels.
\label{fig:N+Mlvlsys}}
\end{figure}

Let $\rho$ denote the density matrix of an $N$ level quantum system (see 
figure \ref{fig:N+Mlvlsys}). The density matrix evolves under the Liouville von Neumann (LVN) equation which takes 
the form
\begin{equation}\label{eq:main}
\dot{\rho} = -i[H(t), \rho] + L(\rho)
\end{equation}where $-i[H, \rho]$ is the unitary evolution of the 
quantum system and $L(\rho)$ is the dissipative part of the evolution. 
The term $L(\rho)$ is linear in $\rho$ and is given by the Lindblad form 
\cite{Lindblad,Alicki86}, i.e.
$$L(\rho) = \sum_{ij}F_{ij}\rho F_{ij}^{\dagger} - \frac{1}{2}
\{F_{ij}^{\dagger}F_{ij}, \rho \} , $$ where
$F_{ij}$ are the Lindblad operators. In this manuscript, we assume
the only relaxation mechanism is spontaneous emission and
therefore we take $F_{ij} = \sqrt{\gamma_{ij}} E_{ij}$ where the
operator $E_{ij} = \ket{i}\bra{j}$ and $\gamma_{ij}$ represents
the rate of spontaneous emission from level $j$ to level $i$. Eq.
(\ref{eq:main}) has the following three well known properties: 1)
${\rm Tr}(\rho)$ remains unity for all time, 2) $\rho$ remains a
Hermitian matrix, and 3) $\rho$ stays positive semi-definite, i.e.
that $\rho$ never develops non-negative eigenvalues.

\noindent The first property follows from \beq {\rm Tr}
(\dot{\rho})={\rm Tr}(-i[H,\rho]) + {\rm Tr}(L\rho) = 0. \eeq The
second property follows from the fact that $ \dot{\rho} =
\dot{\rho}^\dag $ and therefore $\rho(t)=\rho^\dag(t)$. We will
later derive an explicit expression for the evolution of the
spectrum of the density operator under dissipation. The third
property will then be shown as an immediate consequence of this
result.

\subsection{Definitions of Purity}
The density matrix is capable of describing any mixed state in
quantum mechanics, ranging from pure states that are solutions of
the TDSE, to completely incoherent states. There are several
common ways of characterizing how close an arbitrary mixed state
$\rho$ is to a pure state. These measures can be generally termed
purity measures or purities. We use $P(\rho)$ to denote the purity
of the density operator $\rho$.

The most common, and perhaps the simplest measure is
$\rm{Tr}(\rho^2)$ \cite{Messiah58,Bartana,TannorRot,Zeilinger}.
For any density matrix, $0<\rm{Tr}(\rho^2)\leq 1$, with equality
only for a pure state.  Thus, the larger the value of
Tr($\rho^2$), the closer a state is to being pure.  Another useful
measure is the von Neumann entropy, $S_{VN} = -k\ Tr (\rho \ln
\rho)$ \cite{vonNeumann55}.  The von Neumann entropy goes to zero
for a pure state and is greater than zero for any mixed state, and
thus the size of the von Neumann entropy is a measure of the
degree of impurity of a state. Two other measures are the largest
eigenvalue of $\rho$, $|\rho|_\infty$, which goes to 1 for a pure
state and is less than one for a mixed state; and a measure based
on the expansion of the characteristic equation for $\rho$, which
has Tr($\rho^2$) as its leading term, but also takes into account
higher order terms, e.g. Tr($\rho^3$) \cite{Byrd03}.\footnote{
The purity function can be thought of mathematically as a
(partial) ordering over the set of allowed eigenvalues such that
the totally pure state having the spectrum $[1,0,...,0]$ yields
the greatest value of purity and the totally mixed state with
spectrum  $[\frac{1}{N}, \frac{1}{N},...,\frac{1}{N}]$ yields the
lowest. A necessary minimum of structure on the purity ordering is
provided by the concept of majorization \cite{Marshall79,Bhatia97}. 
Let $x$ and $y$ be two $d$-dimensional
real vectors. We use the notation $x^\downarrow$ to indicate the
vector whose entries are the entries of $x$, arranged into
decreasing order, $x_1^\downarrow\ge x_2^\downarrow\ge\cdots \ge
x_d^\downarrow$. We say $x$ is majorized by $y$, written $x\prec
y$, if \beq \sum_{j=1}^k x_j^\downarrow\le\sum_{j=1}^k
y_j^\downarrow, \eeq for $k=1,\cdots,d$, with equality when $k=d$.
Loosely speaking, this definition gives quantitative meaning to
the amount of disorder or mixing in a collection of real numbers.
For example, for any $0\le t\le 1$,
$[\frac{1}{2},\frac{1}{2}]\prec[t,1-t].$
For any $d$ dimensional probability distribution $p$,
$[\frac{1}{d},\cdots,\frac{1}{d}]\prec[p_1,\cdots,p_d].$
Note that there are vectors $x$ and $y$ which are incomparable in
the sense that neither $x\prec y$ nor $y\prec x$ (for example
$x=[0.5,0.25,0.25]$ and $y=[0.4,0.4,0.2]$); majorization therefore
gives only partial ordering. Any reasonable measure of purity
should respect the majorization relation, namely  for two
eigenvalue distributions we should have $P(\lambda') \leq
P(\lambda'')$ if $\lambda'\prec\lambda''$. Such
functions are termed {\it Schur-convex.}}

In general, as is apparent from the above discussion, the entire density 
matrix $\rho$ is not  needed in order to characterize the purity of the 
system; rather, all that is necessary is the set of eigenvalues 
$\lambda$ of $\rho$. All purities can therefore be defined as functions 
solely of the eigenvalues, i.e 
\begin{equation}
P(\rho)=P(\lambda(\rho)).
\end{equation}

We will use the following definition of purity
for the remaining part of the paper.
\begin{definition}
Given the density operator $\rho$, with spectrum $\lambda$, define
its purity $P(\rho)$ as the  largest eigenvalue of $\rho$, i.e.
\beq
P(\rho) = |\rho|_\infty=\lim_{n\to\infty} Tr(\rho^n)^{1\over n}=\lad_1. 
\label{eq:def_purity}
\eeq

\end{definition}
Here $\lad$ is the vector of eigenvalues of $\rho$ arranged in
a decreasing order; for the remainder of this paper the
superscript $\downarrow$ will be assumed every time $\lambda$ is written. 
Although many of the results in this paper
are very general, we choose this measure as it gives simple
answers for the cooling strategies. We will often use $P(\rho)$ or
$P(\lambda)$ to mean the same thing, where it is understood that
$\lambda$ is the spectrum corresponding to $\rho$.

\subsection{\label{sec:Prob}Formulation  of the Control Problem}
The problem we address in this paper is the control of purity content
of a 
quantum dissipative system which evolves under the LVN equation of motion given by eq.~(\ref{eq:main}). The Hamiltonian $H[E]$ depends on 
an externaly controlled laser field $E(t)$ through the dipole coupling term.
Beginning with the system in an initial mixed state it is required to find a control field functionality $E(t)$ that will drive the system through its equations of motion (\ref{eq:main}) to maximal purity, as defined by eq.~(\ref{eq:def_purity}), at  some final time $T$. 

The system evolution equation contains both a Hamiltonian part,
$$\dot{\rho}_H = -i[H[E],\rho],$$
and a dissipative part, given by 
$$\dot{\rho}_D=L(\rho).$$
The Hamiltonian term leads to unitary evolution which does not 
change the spectrum, and the purity depends only on the 
spectrum.  Thus, the dissipative term is required to obtain a purity increase. In \cite{TannorRot}, the control problem was solved completely for the two-level system. In this paper we develop a formalism applicable to general $N$-level systems.
\section{\label{sec:Reform}Reformulation of the Control Problem in Terms of the Spectrum of $\rho$}
\subsection{\label{sec:assump}Simplifying assumptions: complete and instantaneous unitary control}
In this section we develop a general formalism that highlights the cooperative interplay between Hamiltonian and dissipative dynamics. Following 
\cite{TannorRot}, we assume that the action of the control Hamiltonian
can be produced on a time scale fast compared with spontaneous emission. This assumption is well established on physical grounds, since femtosecond laser 
control is now widely available and typical spontaneous emission times are
nanosecond.
In this paper we  make an additional useful simplifying assumption about the dynamics, namely that
the control Hamiltonian $H(t)$ can produce any unitary transformation $U \in SU(N)$ in the 
$N$ level system, i.e. the system of interest is unitarily controllable.
Combining these two assumptions we have that any unitary transformation can be produced on the system in 
negligible time compared to the dissipation.

We use the notation
$$\o(\rho) = \{ U \rho U^\dag | U \in SU(N) \},$$ to denote the orbit 
of $\rho$ under unitary transformations. 
Since $\lambda(U \rho U^{\dag})=\lambda(\rho)$, it is 
obvious that $P(\rho)=P(\lambda)$  is constant along the orbit $\o(\rho)$; however 
$\dot P$ is not: the rate of change of 
the purity due to dissipation is affected by where in $\o(\rho)$ the density matrix
resides.  In other words, due to the 'instantaneous controllability' assumption, unitary controls can instantaneously direct $\rho$ along 
the orbit in order to change $\dot P$ in a controlled manner.

The above dynamical assumptions lead to another very important simplification. 
Since we have assumed that all unitary transformations in $SU(N)$ can be produced 
instantaneously, this includes bringing the density matrix into diagonal form. 
As a result, the different elements of each orbit can be considered redundant,
and the orbit of $\rho$ can be completely represented by its diagonal form, or 'spectrum',
$\lambda(\rho)$.  This suggests reformulating the control problem entirely in terms of the spectrum,
rather than in terms of $\rho$ itself.  
The key step in this reformulation is to replace the equation of motion for $\rho$, eq.~(\ref{eq:main}), with an equation of motion for the spectrum.
We do this in the next
section.  As the purity is a function solely of the spectrum, this equation will allow the
optimization to be performed just on the set of allowed
spectra, significantly reducing the complexity of the problem.  The controls will
enter into this equation in a modified way that gives additional insight into the
interplay of Hamiltonian and dissipative dynamics.

\subsection{\label{sec:Motion}Equations of Motion for the Eigenvalues 
Assuming Fast Unitary Evolution}
Suppose that $\rho$ has a nondegenerate spectrum, and let
$\Lambda$ be its associated diagonal form. Consider two unitary
transformations, $U_1$ and $U_2$. Then both $\rho_1 = U_1 \Lambda
U_1^{\dagger}$ and $\rho_2 = U_2 \Lambda U_2^{\dagger}$  belong to
$\o (\rho)$. However, they do not have the same spectrum after
evolution under the dissipative dynamics. To understand how the
spectrum of the density operator evolves, note that Hamiltonian
dynamics produces no change in the spectrum. Therefore, the change
in the spectrum is solely due to dissipation. After small time
$\delta t$ the initial density operator $\rho$ evolves to

\beq\label{eq:evolve.o} \rho \to \rho+ L(\rho)\delta t, \eeq

If $\Lambda$ represents the diagonalization of the original
density operator $\rho$ ($\rho = U \Lambda U^{\dagger}$, where
$U$ is unitary) then the new density operator can be written as

\beq\label{eq:evolve} \rho \to \rho+ L(\rho)\delta t = 
U(\Lambda + U^\dag L(U\Lambda U^\dag) U \delta t) U^\dag, \eeq

Consider now the change in spectrum under the evolution of 
eq.~(\ref{eq:evolve}). Since $\Lambda$ is diagonal, the spectrum on
the right hand side is, to first order in $\delta t$, just the
diagonal\footnote{This is simply the well known result of first
order perturbation theory which, when applied to a perturbed
Hamiltonian, states that the first order corrections to the
energies are the diagonal elements of the perturbing Hamiltonian
$V$.} 
i.e.
$$ \lambda(t + \delta t) = diag( \Lambda + U^\dag L(U\Lambda 
U^\dag)U \delta t). $$  
Given the matrix $A$, the notation $diag(A)$ represents a
vector whose entries are the diagonal entries of $A$. The rate of
change of eigenvalues is then \beq\label{eq:dotlam}
\dot\lambda=diag(U^\dag L(U\Lambda U^\dag) U) \eeq which is in
general different for different choices  of $U$. Thus by applying
varying unitary transformations $U$ and letting the dissipative
dynamics evolve for some small time $\delta t$ we get different
evolution of the spectrum. The unitary transformation should
therefore be thought of as a control by which the spectrum of the
density matrix can be affected.

\subsection{\label{sec:Canon}Canonical decomposition}
To proceed further, observe that the right hand side of eq.
(\ref{eq:dotlam}) describing the change in the spectrum under
operation of the  Lindbladian is a linear transformation on the
vector of eigenvalues (see appendix \ref{sec:AppCanon})
$$\dot\lambda=M\lambda.$$
To obtain an explicit expression for $M$ first note that for $U=I$
in eq.~(\ref{eq:dotlam}) we have $\dot\lambda=A\lambda$ with $A$ a
$Q$-matrix (columns sum to zero) defined by $A_{ij}=\gamma_{ij}$
for $i\neq j$ and $A_{ii}=-\sum_k\gamma_{ki}$ otherwise. We split
$$A = B + D$$ where $D$ is the diagonal part of $A$ and is all
non-positive whereas $B$ contains all off-diagonal entries and is
all nonnegative. Using these definitions we get for general $U$ in
eq.~(\ref{eq:dotlam}) (for details see appendix
\ref{sec:AppCanon}): 
\beq\label{eq:canonical} \dot\lambda = (
\Theta^{T} B \Theta + \Theta^{T} \circ D ) \lambda , 
\eeq where
$ \Theta_{ij} = |U_{ij}|^2 $, is the Schur product of $U$ with its
complex conjugate. Note that $\Theta$ has the important property
of being  a doubly-stochastic matrix (rows and columns all sum to
unity). The notation $\Theta^{T} \circ D$ denotes the linear
transformation of the diagonal of $D$ (as a vector) under the
action of $\Theta^{T}$. In other words, if $d=diag(D)$, then 
$\Theta^{T} \circ D$ is a
diagonal matrix whose diagonal is $\Theta^{T} d$. Note that in
the special case where $U\in\{P_i\}$ --- the set of permutations --- 
$\Theta=P_i$, $P_i^T \circ D = P_i^T D P_i$
and hence eq.(\ref{eq:canonical}) simplifies to $\dot\lambda=\Theta^T
A\Theta\lambda$. 

Eq. (\ref{eq:canonical}) is one of the central results  of this
paper; it provides a reduced equation of motion for the spectral
evolution under Lindblad dissipation and parametrized by the
unitary control. 
From eq. (\ref{eq:canonical}), it is
straightforward to infer, for example,  that the eigenvalues of the density operator
always remain nonnegative. In order to become negative an eigenvalue must pass through zero. 
If any of the eigenvalues $\lambda_j=0$, however, the only contributions to $\dot \lambda_j$ will be nonnegative since the only nonpositive elements in $M=
\Theta^{T} B \Theta + \Theta^{T} \circ D$ reside on the diagonal. Hence none of the eigenvalues can turn negative.

\subsection{Revised definition of the Control problem}
Having formulated an equation of motion for the spectrum, eq. (\ref{eq:canonical}), we can now redefine the control problem in terms of the spectrum alone.
We seek a control strategy in the form of a time varying unitary-stochastic matrix $\Theta(t)$ which when applied to the spectral equation of motion (\ref{eq:canonical}), will produce maximal purity $P(\lambda)$ at the final time $T$.

One strategy for choosing $\Theta(t)$ is to instantaneously
maximize the purity $P(\lambda)$ at each point in time.
Maximization algorithms that utilize this strategy are termed
'greedy' algorithms and do not in general guarantee obtaining
maximum possible purity at the final time $T$. To calculate the globally
optimal cooling strategy we use the principle of dynamic
programming \cite{Bertsekas87}, as described in the next section.

\section{\label{sec:HJB}Dynamic Programming and the 
Hamilton-Jacobi-Bellman PDE }

We now use the principle of dynamic programming for finding the
optimal $\Theta(t)$ in Eq. (\ref{eq:canonical}). We will develop
the basic ideas through the problem under consideration. Let
$V(\lambda, t)$ denote the maximum achievable purity starting from
initial eigenvalue spectrum $\lambda$ at time $t$ ($T-t$ units of
time remaining). By definition of $V(\lambda, t)$, it is the
maximum achievable purity if $\Theta$ is chosen optimally over the
interval $[t, T]$. Suppose that at time $t$, the spectrum of $\rho(t)$
is $\lambda(t)$ and we make a choice of $\Theta(t)$. The resulting
density operator after time $\delta t$ depends on the choice of
$\Theta(t)$. The choice of $\Theta(t)$ should be such that for the
resulting new spectrum $\lambda(t+ \delta t)$, the return
function, $V(\lambda(t + \delta t), t + \delta t)$ is maximized
and by definition of the optimal return function should be same as
$V(\lambda(t), t)$. By a Taylor series expansion we obtain
\beqa
V(\lambda(t + \delta t), t + \delta t)&= &
V(\lambda(t), t)+  \frac{\partial V(\lambda, t)}{\partial
t}\delta t\nonumber\\&&+ \max_{\Theta} \langle \frac{\partial V(\lambda,
t)}{\partial \lambda}, \delta {\lambda}(\Theta) \rangle.
\eeqa
This then gives the well known Hamilton-Jacobi-Bellman PDE
\begin{equation}
\label{eq:HJB} \frac{d V(\lambda, t)}{dt} = \frac{\partial
V(\lambda, t)}{\partial t} + \max_{\Theta} \< \frac{\partial
V(\lambda, t)}{\partial \lambda}, \dot{\lambda}(\Theta) \> = 0.
\end{equation}
\n Observe that at the final time $T$, the value of the return function
is just the purity of the density operator, i.e. $$ V(\lambda, T)
= P(\lambda). $$
\n If we solve this PDE, together with its final condition,
 we will get the optimal control $\Theta$
as a function of the spectrum $\lambda$ and the time $t$, denoted as
$\Theta = \Theta^{\ast}(\lambda, t)$. In other words, given
the spectrum $\lambda$ of the density operator at time $t$,
the best cooling strategy is to choose $\Theta^{\ast}(\lambda,
t)$. This implies that the control problem is solved not just for
a particular set of initial conditions; rather, it is embedded in
a wider problem and a solution is sought simultaneously for all
possible initial conditions.

\n In equation (\ref{eq:HJB}), the term $\frac{\partial V(\lambda,
t)}{\partial t}$ has no dependence on $\Theta$, therefore $$
\Theta^{\ast}(\lambda, t) = \arg\max_{\Theta}\< \frac{\partial
V(\lambda, t)}{\partial \lambda}, \dot{\lambda}(\Theta) \>. $$
Substituting for $\dot{\lambda}(\Theta)$ from Eq.
(\ref{eq:canonical}), yields
\beq\label{eq:optTh}
\Theta^{\ast}(\lambda, t) = \arg\max_{\Theta}\<
\frac{\partial V(\lambda, t)}{\partial \lambda}, (\Theta^T B \Theta
+ \Theta \circ D) \lambda \>.
\eeq

\n Thus the problem reduces to finding the optimal control
$\Theta^{\ast}(\lambda, t)$ that maximizes the expression 
\beq
\label{eq:F_Th}
F(\Theta)\equiv\mu^T\left(\Theta^TB\Theta+\Theta^T\circ
D\right)\lambda. 
\eeq

\n where the vector $\mu$ is defined as $\mu_j={\partial
V\over\partial \lambda_j}$ (Although $\mu$ is a function of
$\lambda$ and $t$, we just use $\mu$ and keep in mind that the
dependence is implied). Note that a priori $V(\lambda, t)$ and hence
$\mu$ are not known. However if we can make a guess at the optimal
control strategy (which depends on $\lambda$ and $t$) and use this
optimal strategy to integrate the equation of motion of the system
evolution to obtain $V(\lambda, t)$ and hence $\mu$, then we can
verify if the optimal control $\Theta$ and the corresponding $\mu$
satisfy equation (\ref{eq:optTh}). We illustrate this by finding
optimal cooling strategies for a 3-level Lambda system.

\n The following properties of equation (\ref{eq:F_Th}) will be used
subsequently. $\Theta$ being a double stochastic matrix implies
that $[1,1,...1]\Theta=[1,1,...1]$. Furthermore $[1,1,...1](B + D)
= 0$ and therefore $F(\Theta)$ vanishes for $\mu^T=[1,1,...,1]$.
The elements of $\mu$ can therefore be shifted by a constant
amount to make a specific component of $\mu$ vanish without influencing 
the value of $F(\Theta)$.

\section{\label{sec:3lvlAppl}Solution of the optimal control problem for 
the 3 level System}
\subsection{Preliminaries}
\begin{figure}[hbt]
\centerline{\epsfig{file=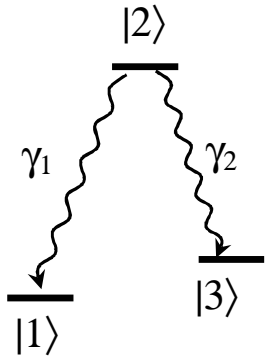,clip=,width=2in}}
\caption[3lvl]{A 3 level Lambda system with spontaneous emission rate 
from level $2$ to $1$ given by $\gamma_1$ and spontaneous emission rate 
from $2$ to $3$ given by $\gamma_2$.
\label{fig:3lvl}}
\end{figure}
Consider a three level Lambda system depicted in figure \ref{fig:3lvl}.
The
 excited state spontaneously decays into the stable ground states 
$\ket{1}$ and $\ket{3}$ at rate $\gamma_1$ and $\gamma_2$ respectively.
 We will assume without loss of generality that $\gamma_1 \geq \gamma_2$.

The evolution of the density matrix of the three level Lambda system is
given by
\beqa\label{eq:main.1}
\dot{\rho} = -i[H(t), \rho]& +& \gamma_1 ( E_1 \rho E_1^{\dagger} 
-\frac{1}{2}\{E_1^{\dagger}E_1, \rho \})\nonumber\\& +& \gamma_2 ( E_2 
\rho E_2^{\dagger} -\frac{1}{2}\{E_2^{\dagger}E_2, \rho \}),
\eeqa
where $E_1 = \ket{1}\bra{2}$ and $E_2 = \ket{3}\bra{2}$.
The equation of motion for the spectrum of the density matrix is then (\ref{eq:canonical})
with $A$, $B$ and $D$ given by
$$A = \left [ \begin{array}{ccc} 0 & \gamma_1 & 0 \\ 0 & 
-(\gamma_1+\gamma_2) & 0 \\ 0 & \gamma_2 & 0 \end{array} \right];$$
$$
B = \left [ \begin{array}{ccc} 0 & \gamma_1 & 0 \\ 0 & 0 & 0 \\ 0 &
\gamma_2 & 0 \end{array} \right];\ \ D = \left [ \begin{array}{ccc} 0 
& 0 & 0 \\ 0 & -(\gamma_1 + \gamma_2) & 0 \\ 0 & 0 & 0 \end{array} 
\right].$$
The objective is to maximize the purity at time $T$, $P(T)$, as measured by the largest
eigenvalue of $\rho$ (Definition 1).
\subsection{The optimal strategy: Keep $\rho$ diagonal and ordered}
\begin{figure}[ht]
\centerline{\epsfig{file=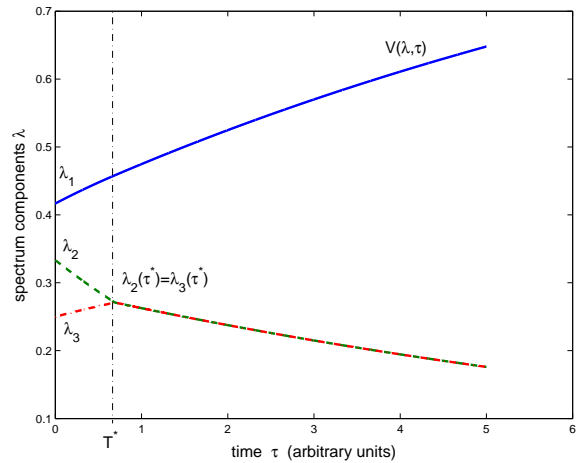,clip=,width=3in}}
\caption[3lvlstrat]{The eigenvalue evolution under the optimal
cooling strategy for the three level $\Lambda$ system.
\label{fig:3lvlstrat}} \end{figure}

Given the equation of motion defined by eq. \ref{eq:main.1} and the objective defined by
Definition 1, we have the following theorem:
\begin{theorem}{\bf The optimal cooling strategy}
{\rm For the 3-level system described above (labelled as in fig. \ref{fig:3lvl}) if any unitary
transformation $U \in SU(3)$ can be produced in arbitrarily small
time, then the optimal cooling strategy is to keep the density
operator $\rho(t)$ diagonal for all times (produce no coherences)
and ordered i.e.  $\rho_{11}(t)\ge\rho_{22}(t)\ge\rho_{33}(t)
$.}
\end{theorem}
\n The optimal control strategy has the following alternate
description. Throughout the cooling process, we keep the largest
eigenvalue in the eigenstate $\ket{1}$, the next largest in state
$\ket{2}$ and finally the smallest in state $\ket{3}$. As the
population in state $\ket{2}$ decays spontaneously to state
$\ket{1}$ and $\ket{3}$, after some time $\tau^{\ast}$, the population
of states $\ket{2}$ and $\ket{3}$ will become equal. From that
point onwards, we always maintain the population of states
$\ket{2}$ and $\ket{3}$, equal (see figure \ref{fig:3lvlstrat}).
We will refer to this strategy as ``greedy'' since it maximizes the rate of increase of the objective at each point in time.

To prove optimality of the
above strategy we proceed as follows. We first compute $V(\lambda,
t)$ for the proposed strategy and then show that it satisfies the
HJB equation maximized over all unitary
transformations. Following the convention that the elements of the vector
$\lambda$ are arranged in decreasing order, this amounts to
showing that 
\beqa
I&=&\argmax_{ \Theta\in \{|U|^2 \ |\ U\in SU(N)\} } 
F(\Theta)\nonumber\\
&=&\argmax_{ \Theta\in \{|U|^2 \ |\ U\in SU(N)\}
}\mu^T\left(\Theta^TB\Theta+\Theta^T\circ D\right)\lambda, 
\eeqa
where $I$ is the identity operator. This implies that the eigenvalues
should be continuously maintained in their ordered arrangement.  Note
that despite the simplicity of this result, in general the continuous 
intervention of a control field is required in order that this condition
be fulfilled.

\subsection{The Return Function for the  Ordered Diagonal Strategy}
\n  We now evaluate the return function for the putative optimal strategy.
Let $\tau = T-t$ denote the remaining
time for cooling. According to the strategy proposed above, two
evolution regimes exist depending on whether
$\tau \leq \tau^{\ast}$ or $\tau > \tau^{\ast}$, where
$\tau^{\ast}$ is the critical time required for $\lambda_2$ and
$\lambda_3$ to come to equilibrium.

\n In the case where $\tau \leq \tau^{\ast}$, under the
proposed strategy the evolution equations of the system take the
form \beqa
\dot\lambda_1 &=& \gamma_1 \lambda_2 \\
\dot\lambda_2 &=& -(\gamma_1 + \gamma_2) \lambda_2 \\
\dot\lambda_3 &=& \gamma_2 \lambda_2 \eeqa and therefore \beq
V(\lambda, T-\tau) = \lambda_1 + \frac{\gamma_1}{\gamma_1 +
\gamma_2} \lambda_2 \left( 1 - e^{-(\gamma_1 +
\gamma_2)\tau}\right);\ \ \tau < \tau^{\ast}. \eeq By definition
$\lambda_2(\tau^{\ast}) =\lambda_3(\tau^{\ast})$ and
$\lambda_2(\tau^{\ast})=\lambda_2e^{-(\gamma_1+\gamma_2)\tau^\ast}$.
Using these equalities, the following explicit forms for
$\lambda_2(\tau^\ast)$ and $\tau^{\ast}$ can be computed \beqa
\lambda_2(\tau^{\ast}) &=& \frac{\gamma_2\lambda_2+(\gamma_1 + 
\gamma_2)\lambda_3}{\gamma_1 + 2 \gamma_2}.\nonumber\\
\tau^{\ast} &=&
- \frac{1}{\gamma_1 + \gamma_2} \log\left(
\frac
{\lambda_2\gamma_2+\lambda_3(\gamma_1+\gamma_2)}
{\lambda_2(\gamma_1 + 2 \gamma_2)}
\right).
\eeqa
After this point in time, under the ordered diagonal policy, the 
populations of states $\ket{2}$ and $\ket{3}$ are maintained at 
equilibrium such that $\lambda_2(\tau) = \lambda_3(\tau) =
\frac{1}{2} (1 - \lambda_1(\tau))$. The system dynamics therefore takes 
the form
$$
\dot\lambda_1 = -\frac{\gamma_1}{2} (1 - \lambda_1),
$$
from which the return function for the regime $\tau >
\tau^{\ast}$, can be explicitly computed. \beq V(\lambda, T- \tau)
=1 - 2 \lambda_2(\tau^{\ast})e^{-\frac{\gamma_1}{2}(\tau-
\tau^{\ast})} ;
 \ \ \tau > \tau^{\ast}.
\eeq

\n As the return function enters the HJB equations only through its 
derivatives $\mu=\frac{\partial V}{\partial \lambda}$, we proceed to 
compute these derivatives explicitly for use in the next section.\\
For $\tau < \tau^{\ast}$, we have
\begin{eqnarray}\label{eq:mus1}
\mu_1 &=& 1 \nonumber\\
\mu_2  &=& \frac{\gamma_1}{\gamma_1 + \gamma_2}\left(1 -e^{-(\gamma_1 
+ \gamma_2)\tau}\right) \nonumber\\
\mu_3&=& 0
\end{eqnarray}
and for $\tau > \tau^{\ast}$, we have
\begin{eqnarray}\label{eq:mus2}
\mu_1 &=& 0 \nonumber\\
\mu_2 &=& - \frac{2\gamma_2\lambda_2+\gamma_1\lambda_3}{\lambda_2(\gamma_1 + 2 
\gamma_2)}e^{-\frac{\gamma_1}{2}(\tau-\tau^{\ast})} \nonumber\\
\mu_3 &=& -e^{-\frac{\gamma_1}{2}(\tau-\tau^{\ast})}
\end{eqnarray}
Note that in both regimes $\mu_1\ge\mu_2\ge\mu_3$, a property that
will be used below.\footnote{In order to prove this statement for
$\tau< \tau^{\ast}$ note that in this regime
$\lambda_3(\tau)\le\lambda_2(\tau)$, which implies
$\frac{\lambda_2}{\gamma_1+\gamma_2}\left(
(\gamma_1+2\gamma_2)e^{-(\gamma_1+\gamma_2)\tau}-\gamma_2\right)\ge\lambda_3$.}
Also note that the $\mu$'s are continuous at $\tau=\tau^{\ast}$ up to a
constant shift (see remark at the end of section \ref{sec:HJB}).
\subsection{Proof that the Return Function for the Ordered Diagonal 
Strategy Satisfies HJB}
We proceed to calculate $\argmax F(\Theta)$ for $F(\Theta)$ given by
eq. (\ref{eq:F_Th}) and $\mu$ given by eq. (\ref{eq:mus1}) and (\ref{eq:mus2}).  We show that
$\Theta^{\ast}=I$ and hence the ordered diagonal strategy satisfies the
HJB equation, proving that this strategy is globally optimal.

\paragraph{Absence  of ground state coherences in the ordered diagonal solution} 
We first prove that the optimal transformation $\Theta$ in
equation (\ref{eq:F_Th}) has the property that
$\Theta_{13}=\Theta_{31}=0$, namely that the ground state
coherences vanish throughout the evolution of the optimal
trajectory. Suppose $\Theta_{13} \neq 0$ and $\Theta_{31}\neq 0$
and say $\Theta_{31} \geq \Theta_{13}$. From equation (\ref{eq:F_Th}) we have 
\beq
\begin{split}
 F(\Theta)=&\left[\gamma_1(\mu_1 \Theta_{11} + \mu_3
\Theta_{13}) + \gamma_2 ( \mu_1 \Theta_{31} + \mu_3 \Theta_{33} )
\right]\\
&\times \left[\lambda_1 \Theta_{21} + \lambda_2 \Theta_{22} + \lambda_3
\Theta_{23}\right]\\
&- (\gamma_1 + \gamma_2) [\mu_1 \lambda_1 \Theta_{21}
+ \mu_3 \lambda_3 \Theta_{23}],
\end{split}
\eeq
where we have chosen $\mu_2 = 0$
and hence $\mu_1 \geq 0\geq \mu_3$. Let $\Delta =
\Theta_{13}$. Observe that in the above equation we can increase $\Theta_{11}$ and $\Theta_{33}$ by an amount $\Delta$ and decrease
$\Theta_{13}$ and $\Theta_{31}$ by $\Delta$, to generate a new
doubly stochastic matrix which gives a larger value of $F(\Theta)$
(this follows from the relations $\gamma_1 \geq \gamma_2$ and $\mu_1\geq 0\geq \mu_3$).
Hence we assume $\Theta_{13} = 0$. Let $\Delta_1$ be the new value
of $\Theta_{31}$. Now if we increase $\Theta_{11}$ and
$\Theta_{32}$ by $\Delta_1$ and decrease $\Theta_{31}$ and
$\Theta_{12}$ by the same amount we get a new doubly stochastic
matrix which gives a larger value of $F(\Theta)$. Hence we
need to maximize $F(\Theta)$ only over those doubly stochastic matrices
for which $\Theta_{13}=\Theta_{31}=0$.

\paragraph{Dependence of $F(\Theta)$ on the remaining parameters in 
$\Theta$}
As the rows and columns of $\Theta$ must sum to unity there remain only 
two degrees of freedom in the components of $\Theta$. Therefore, we can 
write $F(\Theta)$ as a function of only two of its components:
\beqa
F(\Theta)&=&F(\Theta_{21},\Theta_{23})\nonumber\\
&=&
\left[\gamma_1\mu_1(1-\Theta_{21})+\gamma_2\mu_3(1-\Theta_{23})\right]\nonumber\\
&&\times\left[\lambda_2+\Theta_{21}(\lambda_1-\lambda_2)+\Theta_{23}(\lambda_3-\lambda_2)\right]\nonumber\\
&&-(\gamma_1+\gamma_2)\left[\mu_1\Theta_{21}\lambda_1+\mu_3\Theta_{23}\lambda_3\right].
\eeqa
It is now required to find the maximum of $F(\Theta_{21},\Theta_{23})$ 
on the triangular domain $0\leq\Theta_{21}\leq 1$, $\quad 0\leq\Theta_{23}\leq 1$, $\Theta_{21}+\Theta_{23}\leq1$.

\paragraph{The maximum cannot lie at an interior point.}
Suppose $F$ has a maximum in the interior, then the Hessian of $F$ at
that point must be negative definite. We proceed to show that the
Hessian $G_{ij}\equiv \frac{\partial^2 F}{\partial
\Theta_{2i}\partial\Theta_{2j}}$, with $i,j=\{1,3\}$, is {\it
not} negative definite anywhere and therefore the maximum  must
reside on the boundary. Computing the components of $G$ we find
\beqa
G_{11}&=&-2(\lambda_1-\lambda_2)\mu_1\gamma_1\nonumber\\
G_{33}&=&-2(\lambda_3-\lambda_2)\mu_3\gamma_2\nonumber\\
G_{31}&=&G_{13}=-\mu_1(\lambda_3-\lambda_2)\gamma_1-\mu_3(\lambda_1-
\lambda_2)\gamma_2.
\eeqa Denoting $a\equiv\gamma_1\mu_1(\lambda_3-\lambda_2)$ and
$b\equiv\gamma_2\mu_3(\lambda_1-\lambda_2)$, the determinant of $G$ is
$4ab-(a+b)^2=-(a-b)^2\leq 0$ such that one of the eigenvalues of
$G$ is non negative  and therefore $G$ is not negative definite.

\paragraph{The maximum point is $(\Theta_{21},\Theta_{23})=(0,0).$}
As the maximum does not reside in the interior of the triangular domain 
it must lie on one
of the edges $[(0,0),(0,1)]$, $[(0,0),(1,0)]$ or $[(0,1),(1,0)]$.

It can be shown by checking the first and second derivatives along
the edge $[(0,1),(1,0)]$ that the maximum along that interval
lies at the end point $(\Theta_{21},\Theta_{23})=(0,1)$. We now
check the remaining two edges. As $G_{11}$ and $G_{33}$ are
both non positive it follows that $F(\Theta)$ is concave in both
the $\Theta_{21}$ and $\Theta_{23}$ directions. Therefore, if in
addition the slope at the point $(0,0)$ is negative in both
directions, this establishes the existence of a maximum at that
point. We proceed to show that indeed the slopes are non negative:
\beqa \left.\frac{\partial F}{\partial \Theta_{21}}\right|_{(0,0)}
&=&(\lambda_1-\lambda_2)[\mu_3\gamma_2+\mu_1\gamma_1]+\lambda_2[-\mu_1
\gamma_1]\nonumber\\
&&-(\gamma_1+\gamma_2)\mu_1\lambda_1 \leq 0\\
\left.\frac{\partial F}{\partial \Theta_{23}}\right|_{(0,0)}
&=&(\lambda_3-\lambda_2)[\mu_3\gamma_2+\mu_1\gamma_1]+\lambda_2[-\mu_3
\gamma_3]\nonumber\\
&&-(\gamma_1+\gamma_2)\mu_3\lambda_3 \leq 0.
\eeqa
The first expression follows from the fact that $\gamma_1\ge\gamma_2$, $\mu_1\ge 0\ge\mu_3$ and $\lambda_1\ge\lambda_2$.
The second expression can be proved by inserting the explicit forms for 
$\mu$, eq.~(\ref{eq:mus1}) and (\ref{eq:mus2}), for the two regimes of $T-t$.

\section{\label{sec:Conc}Discussion and Conclusions}

We have presented a general framework for calculating optimal purity
increasing strategies in $N$ level dissipative systems under the assumption of 
complete and instantaneous unitary control. In so doing, we derived  a 
reduced equation of motion for the spectral evolution under dissipation 
and parametrized by the unitary control. The Hamilton-Jacobi-Bellman 
Theorem was invoked to provide sufficient criteria for global optimality.
This general framework was then explicitly applied to derive and prove 
optimality of the greedy cooling strategy for a three level $\Lambda$ system.

In future work we intend to apply this methodology to obtain
explicit optimal cooling strategies for general $N+M$ level systems
comprised of $M$ excited states coupled to $N$ ground states. 
One is tempted, by extrapolation from the present results, to assume
that the greedy algorithm should be optimal in general and hence that
coherences do not play a role in the optimal cooling strategy for 
$N > 3$.  However, preliminary numerical results based on dynamical
programming show that the greedy algorithm is in general not optimal
in these systems.  Rather, a strategy based on 
``delayed gratification'' is superior to the greedy strategy, and coherences
play a small but finite role in these larger systems.  This will be the 
subject of a future paper.

\appendix
\section{\label{sec:AppCanon}Derivation of the 'Canonical form'}

We wish to show that eq. (\ref{eq:dotlam}) is a linear transformation of 
the form $$\dot\lambda=M\lambda.$$ Recall first that
\beq
\begin{split}
U^\dagger& L(U\Lambda U^\dagger)U\nonumber\\
&=\sum_{ij}\gamma_{ij}U^\dagger\left[\ket{i}\bra{j}U\Lambda=
U^\dagger\ket{j}\bra{i}-\frac{1}{2}\left\{\ket{j}\bra{j},U\Lambda 
U^\dagger\right\}\right]U\nonumber\\
& =\sum_{ij}\gamma_{ij}\left[U^\dagger\ket{i}\bra{j}U\Lambda 
U^\dagger\ket{j}\bra{i}U-\frac{1}{2}\left\{U^\dagger\ket{j}\bra{j}U,\Lambda\right\}\right]
\end{split}
\eeq
Expanding eq. (\ref{eq:dotlam}) and rewriting it in component form we have
\beqa
\dot\lambda_k&=&
\sum_{ij}\gamma_{ij}\left[\sum_s U^\dagger_{ki}U_{js}\lambda_s 
U^\dagger_{sj}U_{ik}\right.\nonumber\\
&&\quad\quad\quad\left.-\frac{1}{2}\{U^\dagger_{kj}U_{jk}\lambda_k+
\lambda_kU^\dagger_{kj}U_{jk}\}\right]\nonumber\\
&=&\sum_s\left[\sum_{ij}\Theta^T_{ki}\gamma_{ij}\Theta_{js}-\sum_j
\Theta^T_{kj}\sum_i\gamma_{ij}\delta_{ks}\right]\lambda_s\nonumber\\
&\equiv&\sum_s M_{ks}\lambda_s,
\eeqa
with
$$
M\equiv \Theta^{T} B \Theta + \Theta^{T} \circ D,
$$
and with the definitions of $\Theta$, $B$, $D$ and the operation 
$\Theta\circ D$ as provided in the main text. Rewriting the above in 
vector format we have  precisely eq. (\ref{eq:canonical})
$$
\dot\lambda=(\Theta^{T} B \Theta + \Theta^{T} \circ D)\lambda.
$$

\end{document}